\documentclass{PoS}

\PoS{PoS(LAT2005)076}

\title{Combining Quark and Link Smearing to Improve Extended Baryon Operators}

\ShortTitle{Smearing Extended Baryon Operators}

\author{Subhasish Basak, Ikuro Sato, Stephen Wallace\\
        Department of Physics, University of Maryland, College Park, MD 20742, USA\\
        E-mail: \email{sbasak@glue.umd.edu},
	\email{ikuro@glue.umd.edu},
	\email{stevewal@physics.umd.edu}}
\author{Robert Edwards, David Richards\\
        Thomas Jefferson National Accelerator Facility, Newport News, VA 23606, USA\\
        E-mail: \email{edwards@jlab.org},
	\email{dgr@jlab.org}}
\author{George T. Fleming\\
        Sloane Physics Laboratory, Yale University, New Haven, CT 06520, USA\\
        E-mail: \email{george.fleming@yale.edu}}
\author{Urs M. Heller\\
        American Physical Society, Ridge, NY 11961-9000, USA\\
        E-mail: \email{heller@csit.fsu.edu}}
\author{\speaker{Adam Lichtl} and Colin Morningstar\\
        Department of Physics, Carnegie Mellon University, Pittsburgh, PA 15213, USA\\
        E-mail: \email{alichtl@andrew.cmu.edu},
	\email{colin\_morningstar@cmu.edu}}

\abstract{The effects of Gaussian quark-field smearing and 
analytic stout-link smearing on the correlations of gauge-invariant 
extended baryon operators are studied.  Gaussian quark-field
smearing substantially reduces contributions from the short wavelength modes
of the theory, while stout-link smearing significantly reduces the noise from 
the stochastic evaluations.  The use of gauge-link smearing is shown to
be crucial for baryon operators constructed of covariantly-displaced
quark fields.  Preferred smearing parameters are determined for a 
lattice spacing $a_s\sim0.1$~fm.}

\FullConference{XXIIIrd International Symposium on Lattice Field Theory\\
                 25-30 July 2005\\
                 Trinity College, Dublin, Ireland}

\begin{document}

\section{Introduction}

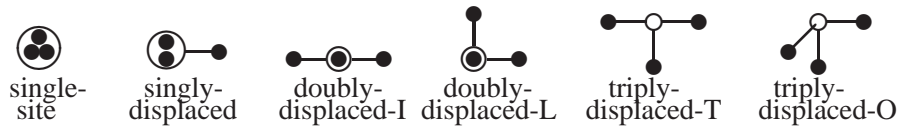
\begin{figure}[b]
\centerline{
\raisebox{0mm}{\setlength{\unitlength}{1mm}
\thicklines
\begin{picture}(16,10)
\put(8,6.5){\circle{6}}
\put(7,6){\circle*{2}}
\put(9,6){\circle*{2}}
\put(8,8){\circle*{2}}
\put(4,0){single-}
\put(5,-3){site}
\end{picture}}
\raisebox{0mm}{\setlength{\unitlength}{1mm}
\thicklines
\begin{picture}(16,10)
\put(7,6.2){\circle{5}}
\put(7,5){\circle*{2}}
\put(7,7.3){\circle*{2}}
\put(14,6){\circle*{2}}
\put(9.5,6){\line(1,0){4}}
\put(4,0){singly-}
\put(2,-3){displaced}
\end{picture}}
\raisebox{0mm}{\setlength{\unitlength}{1mm}
\thicklines
\begin{picture}(20,8)
\put(12,5){\circle{3}}
\put(12,5){\circle*{2}}
\put(6,5){\circle*{2}}
\put(18,5){\circle*{2}}
\put(6,5){\line(1,0){4.2}}
\put(18,5){\line(-1,0){4.2}}
\put(6,0){doubly-}
\put(4,-3){displaced-I}
\end{picture}}
\raisebox{0mm}{\setlength{\unitlength}{1mm}
\thicklines
\begin{picture}(20,13)
\put(8,5){\circle{3}}
\put(8,5){\circle*{2}}
\put(8,11){\circle*{2}}
\put(14,5){\circle*{2}}
\put(14,5){\line(-1,0){4.2}}
\put(8,11){\line(0,-1){4.2}}
\put(4,0){doubly-}
\put(1,-3){displaced-L}
\end{picture}}
\raisebox{0mm}{\setlength{\unitlength}{1mm}
\thicklines
\begin{picture}(20,12)
\put(10,10){\circle{2}}
\put(4,10){\circle*{2}}
\put(16,10){\circle*{2}}
\put(10,4){\circle*{2}}
\put(4,10){\line(1,0){5}}
\put(16,10){\line(-1,0){5}}
\put(10,4){\line(0,1){5}}
\put(4,0){triply-}
\put(1,-3){displaced-T}
\end{picture}}
\raisebox{0mm}{\setlength{\unitlength}{1mm}
\thicklines
\begin{picture}(20,12)
\put(10,10){\circle{2}}
\put(6,6){\circle*{2}}
\put(16,10){\circle*{2}}
\put(10,4){\circle*{2}}
\put(6,6){\line(1,1){3.6}}
\put(16,10){\line(-1,0){5}}
\put(10,4){\line(0,1){5}}
\put(4,0){triply-}
\put(2,-3){displaced-O}
\end{picture}}  }
\vspace*{8pt}
\caption{The spatial arrangements of the extended three-quark baryon
operators. Smeared quark-fields are
shown by solid circles, line segments indicate
gauge-covariant displacements, and each hollow circle indicates the location
of a Levi-Civita color coupling.  For simplicity, all displacements
have the same length in an operator.  Results presented here used
displacement lengths of $3a_s$ ($\sim 0.3 \mbox{ fm}$).
\label{fig:operators}}
\end{figure}

One goal of the Lattice Hadron Physics Collaboration (LHPC) is to 
calculate the low-lying hadron spectrum in QCD\cite{bib:baryons}.  Determining
the spectrum requires extracting excited-state energies from our Monte
Carlo computations, necessitating the evaluation of correlation matrices
of sets of operators.  To extract such energies, operators which couple
strongly to the low-lying states of interest and weakly to the high-lying states must
be used.  The need for extended three-quark operators to capture both the radial and 
orbital structures of baryons has been emphasized and described in 
Ref.~\cite{bib:baryons} (see Fig.~\ref{fig:operators}). 

For single-site (local) hadron operators, it is well known that the use
of spatially-smeared quark fields is crucial.  For extended baryon operators,
one expects quark-field smearing to be equally important, but the relevance
and interplay of link-field smearing is less well known. Thus, we decided that
a systematic study of both quark-field and link-variable smearing was warranted.

\section{The Smearing Procedures}
Damping out couplings to the short-wavelength, high-momentum modes is
the crucial feature which any effective smearing 
prescription\cite{bib:wuppertal,bib:jacobi} must have.  Gaussian suppression of
the high-momentum modes is perhaps the simplest method one can use.  Since a Gaussian in momentum
space remains a Gaussian in coordinate space, we decided to employ a gauge-covariant smearing 
scheme\cite{bib:gaussiansmear} in which the smeared quark field is defined at a given site as
a Gaussian-weighted average of the surrounding sites on the same time-slice:
\begin{equation}\widetilde{\Psi}(\mathbf{x})\sim\int d^3r\ e^{-\mathbf{r}^2/(4\sigma_s^2)}
\ \Psi(\mathbf{x}+\mathbf{r})\sim e^{+\sigma_s^2\Delta/4}\ \Psi(\mathbf{x}).
\end{equation}
In practice, this expression must be approximated by
\begin{eqnarray}
 \tilde{\Psi}(x)&=&\left(1+\frac{\sigma_s^2}{4n_\sigma}
 \Delta\right)^{n_\sigma}\Psi(x),\\
\Delta\Psi(x)&=&\sum_{k=\pm1,\pm2,\pm3}\left(U_k(x)\Psi(x+\hat{k})-\Psi(x)\right),
\end{eqnarray}
where $\Delta$ denotes the three-dimensional gauge-covariant Laplacian.
The two parameters to tune in this smearing procedure are the smearing radius $\sigma_s$ and 
the integer number of iterations $n_{\sigma}$.

APE smearing\cite{bib:apesmear} is the most commonly-used means of smearing the 
gauge-field link-variables.  However, to avoid the abrupt projection back onto the gauge
group needed by this prescription, we decided instead to use the analytic smearing scheme
known as stout-link smearing\cite{bib:stout} defined by
\begin{eqnarray}
  U&\to& U^{(1)}\to U^{(2)}\to \cdots \to U^{(n_\rho)},\\ 
 U^{(n+1)}_k(x)&=&\exp\left(i\rho\Theta^{(n)}_\mu(x)\right)U^{(n)}_k(x),\\
\Theta_k(x)&=&\frac{i}{2}\left(\Omega^\dagger_k(x)-\Omega_k(x)\right)-\frac{i}{2N}\mbox{Tr}\left(\Omega^\dagger_k(x)-\Omega_k(x)\right)\\
\Omega_k(x)&=&C_k(x) U^\dagger_k(x)\qquad\mbox{(no summation over $k$)}\\
C_k(x)&=&\displaystyle\sum_{i\neq k}\left(U_i(x)U_k(x+\hat{\imath})U^\dagger_i(x+\hat{k}) + U^\dagger_i(x-\hat{\imath})U_k(x-\hat{\imath})U_i(x-\hat{\imath}+\hat{k}) \right).
\end{eqnarray}
Only the spatial links are smeared, and $C_k(x)$ is a sum of spatial staples.
The two parameters to tune in this smearing procedure are the staple weight $\rho$ 
and the integer number of iterations $n_\rho$.  The quark-field
and link-field smearing schemes preserve the gauge invariance of hadron operators.

\section{Systematic Study of the Smearing Procedures}

\begin{figure}[b]
\includegraphics[width=6.0in, bb=0 0 567 207]{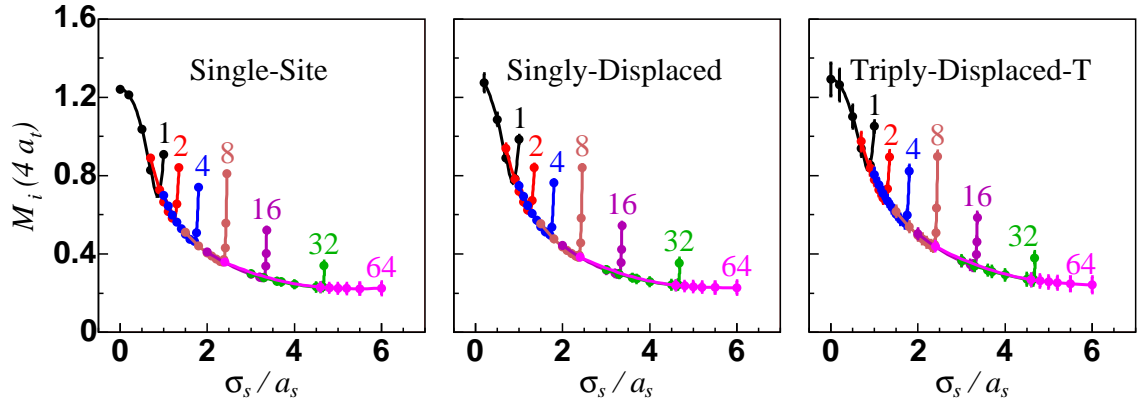}
\caption{$M_i(4a_t)$ for the operators $O_{SS},\ O_{SD},\ O_{TDT}$
against smearing radius $\sigma_s$ for $n_\sigma=1,2,4,8,16,32,64$.  The gauge field
is smeared using $n_\rho=16$ and $n_\rho\rho=2.5$.
Results are based on 50 quenched configurations on a $12^3\times 48$
anisotropic lattice using the Wilson action with $a_s \sim 0.1$ fm
and $a_s/a_t \sim 3.0$.  The quark mass is such that the mass of the pion 
is approximately 700 MeV. 
\label{fig:qtune}}
\end{figure}

In order to study the effects of the smearing procedures on the correlations of
extended baryon operators, we first focused on three particular nucleon operators:
a single-site operator $O_{SS}$ in the $G_{1g}$ irreducible representation of the cubic
point group, a singly-displaced operator $O_{SD}$ with a particular choice of each Dirac
index, and a triply-displaced-T operator $O_{TDT}$ with a specific choice of each
Dirac index (see Fig.~\ref{fig:operators}).

As usual, the effective mass associated with the correlation function 
$C_{ii}(t)=\langle O_i(t)\overline{O}_i(0)\rangle$ is defined by
$M_i(t)=\ln\left(C_{ii}(t)/C_{ii}(t+a_t)\right)$.
To compare the effectiveness of different values of the quark-field smearing parameters,
we compared the effective mass $M_i(t=4a_t)$ for each of the three operators
at a particular temporal separation $t=4a_t$.  Results using 50 quenched 
configurations on a $12^3\times 48$ anisotropic lattice using the Wilson action 
with $a_s \sim 0.1$ fm and $a_s/a_t \sim 3.0$ are shown in Fig.~\ref{fig:qtune}.  
Without gauge-link smearing, the displaced operators were found to be excessively
noisy, making a meaningful comparison impossible.  For this reason, the results 
shown in Fig.~\ref{fig:qtune} include gauge-field smearing with $n_\rho=16$ and 
$n_\rho\rho=2.5$.  One sees that $M_i(t=4a_t)$ is independent of $n_\sigma$
for sufficiently small $\sigma_s$.  For each value of $n_\sigma$, $M_i(t=4a_t)$
first decreases as $\sigma_s$ is increased, until the approximation to a Gaussian 
eventually breaks down, signaled by a sudden rapid rise in $M_i(t=4a_t)$.
This rapid rise occurs at larger values of $\sigma_s$ for larger values of 
$n_\sigma$.

\begin{figure}[t]
\includegraphics[width=6.0in, bb=0 0 567 150]{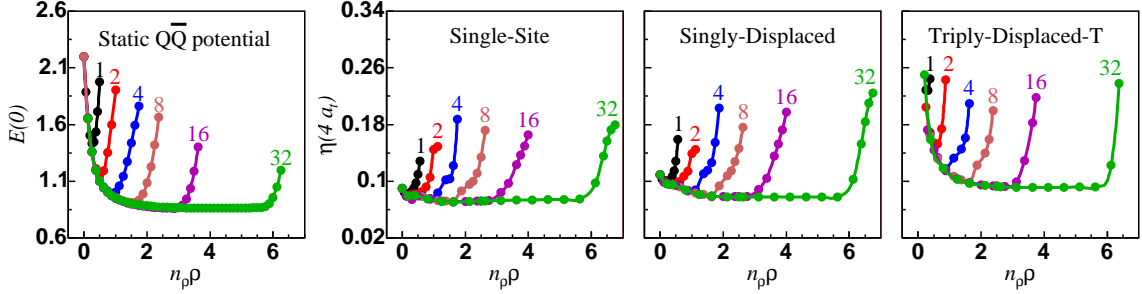}
\caption{
Leftmost plot: the effective mass $E(0)$ for $t=0$
corresponding to the static quark-antiquark potential at spatial separation
 $R = 5a \sim 0.5$ fm against $n_\rho\rho$ for $n_\rho=1,2,4,8,16,32$.
Results are based on 100 configurations on a $16^4$ isotropic lattice using
the Wilson gauge action with $\beta=6.0$.
Right three plots: the relative jackknife error $\eta(4a_t)$ of effective 
masses $M_i(4a_t)$ of the three nucleon operators $O_{SS},\ O_{SD},\ O_{TDT}$
for $n_\sigma=32,\ \sigma_s=4.0$ against $n_\rho\rho$ for $n_\rho=1,2,4,8,16,32$.
Results are based on 50 quenched configurations on a $12^3\times 48$ anisotropic 
lattice using the Wilson action with $a_s \sim 0.1$ fm, $a_s/a_t \sim 3.0$.
\label{fig:ltune}}
\end{figure}

Next, we studied the effect of changing the gauge-field smearing parameters.
First, the effective mass $E(0)$ associated with the static 
quark-antiquark potential at a spatial separation $R=5a_s\sim 0.5$~fm and at a 
particular temporal separation $t=0$ was used to compare the 
effectiveness of different values of $\rho$ and $n_\rho$.  The results are shown 
in the leftmost plot in Fig.~\ref{fig:ltune}.  The behavior is qualitatively
similar to that observed in Fig.~\ref{fig:qtune}.  One sees that the 
$t=0$ effective mass is independent of the product $n_\rho \rho$ 
for sufficiently small values of $n_\rho \rho$.  For each value of $n_\rho$, 
$E(0)$ decreases as $n_\rho\rho$ increases, until a minimum is
reached and a rapid rise occurs.  The onset of the rise occurs at larger
values of $n_\rho\rho$ as $n_\rho$ increases.  Note that $E(0)$ does
not decrease appreciably as $n_\rho\rho$ increases above 2.5.  Hence, 
$n_\rho\rho=2.5$ with $n_\rho=16$ are our preferred values for the link smearing
at lattice spacing $a_s\sim0.1$~fm, based on the static quark-antiquark potential.

Somewhat surprisingly, we found that changing the link-smearing parameters
did not appreciably affect the mean values of the effective masses of our
three nucleon operators.  However, the effect on the variances of the effective
masses was dramatic.   The relative jackknife error $\eta(4a_t)$
of $M(4a_t)$ is shown against $n_\rho\rho$ in the right three
plots in Fig.~\ref{fig:ltune}, and amazingly, this error shows the same 
qualitative behavior as in Fig.~\ref{fig:qtune} and the leftmost plot
in Fig.~\ref{fig:ltune}.  One key point learned here is that the preferred 
link-smearing parameters determined from the static quark-antiquark potential
produce the smallest error in the extended baryon operators.

\begin{figure}[p]
\includegraphics[width=6.0in, bb=0 0 567 559]{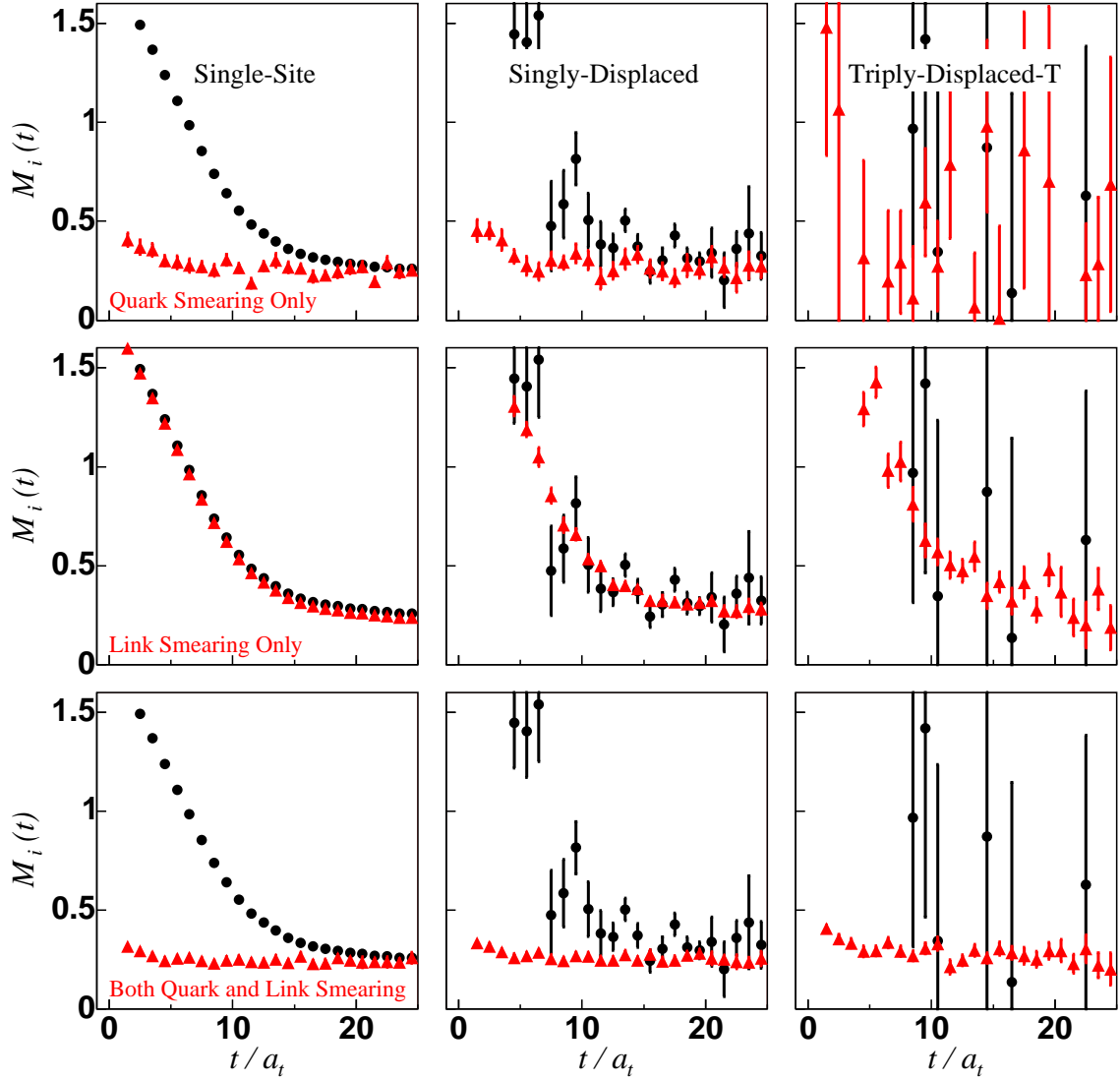}
\caption{Effective masses $M(t)$ for unsmeared (black circles) and smeared 
(red triangles) operators $O_{SS},\ O_{SD},\ O_{TDT}$. 
Top row: only quark-field smearing $n_\sigma=32,\ \sigma_s=4.0$ is used. Middle row: 
only link-variable smearing $n_\rho=16,\ n_\rho\rho=2.5$ is applied.  
Bottom row: both quark and link smearing $n_\sigma=32,\ \sigma_s=4.0, 
\ n_\rho=16,\ n_\rho\rho=2.5$ are used, dramatically improving the signal for all
three operators. Results are based on 50 quenched configurations on a 
$12^3\times 48$ anisotropic lattice using the Wilson action with $a_s \sim 0.1$ fm,
$a_s/a_t \sim 3.0$.\label{fig:meff-smear}}
\end{figure}

\begin{figure}[t]
\includegraphics*[width=6.0in, bb=0 0 567 207]{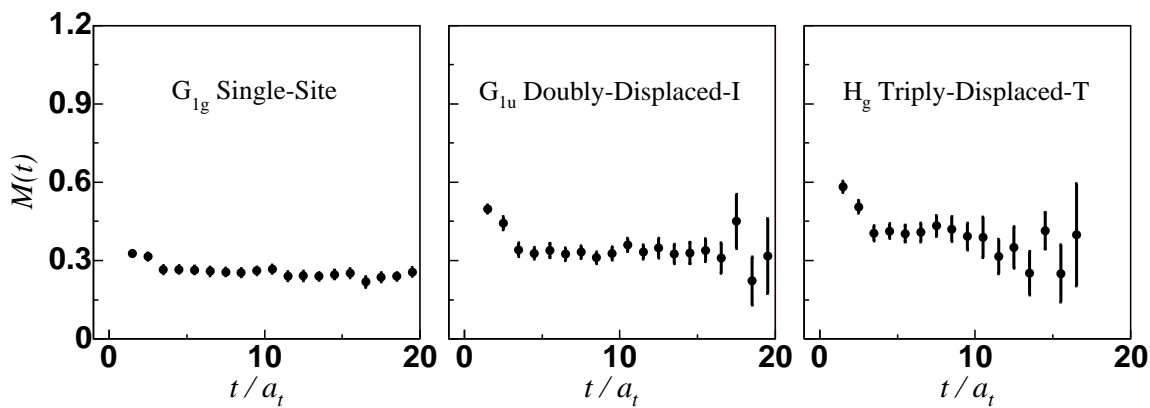}
\caption{Effective masses for three selected nucleon operators:
a single-site operator in the $G_{1g}$ channel (left), a doubly-displaced-I
operator in the $G_{1u}$ channel (center), and a triply-displaced-T
operator in the $H_g$ channel.  The smearing parameters used
were $n_\sigma=32,\ \sigma_s=4.0,\ n_\rho=32,\ n_\rho\rho=2.5$.
Results are based on 25 quenched configurations on a $12^3\times 48$
anisotropic lattice using the Wilson action with $a_s \sim 0.1$ fm and
$a_s/a_t \sim 3.0$.  We have used opposite-parity time-reversed averaging
\protect{\cite{bib:baryons}} to increase statistics.\label{fig:meff-irrep}}
\end{figure}

The effective masses shown in Fig.~\ref{fig:meff-smear} also illustrate
these findings.  The top row shows that applying only quark-field smearing
to the three selected nucleon operators significantly reduces couplings
to higher-lying states, but the displaced operators remain excessively
noisy.  The second row illustrates that including only link-field smearing
substantially reduces the noise, but does not appreciably alter the effective
masses themselves.  The bottom row shows dramatic improvement from reduced
couplings to excited states and dramatically reduced noise when both
quark-field and link-field smearing is applied, especially for the extended
operators. The effectiveness of the smearing schemes used
is further illustrated in Fig.~\ref{fig:meff-irrep}.

\section{Conclusion}
Incorporating both quark-field and link-variable smearing is crucial
for extracting the baryon spectrum using gauge-invariant extended three-quark
operators.  Gaussian quark-field smearing dramatically diminishes couplings to the
short wavelength modes of the theory, whereas stout-link smearing drastically reduces
the noise in operators with displaced quarks.  Preferred smearing parameters
$\sigma_s=4.0,\ n_\sigma=32,\ n_\rho\rho=2.5,\ n_\rho=16$ were found for a 
lattice spacing $a_s\sim 0.1$~fm and were independent of
the baryon operators chosen.  Two issues which remain are to determine
the effects of smearing on the low-lying excited-states, and to determine
the dependence of the preferred smearing parameters on the quark mass.
This work was supported by the U.S.~National Science Foundation
through grants PHY-0099450 and PHY-0300065, and by
the U.S.~Department of Energy under
contracts DE-AC05-84ER40150 and DE-FG02-93ER-40762.

\end{document}